\documentstyle[preprint,aps,epsf]{revtex}

\def\pubnumber{\vbox{\hbox{KUNS
1428}}}

 \def\scri{\hbox{${\cal J}$\kern -.645em {\raise
      .57ex\hbox{$\scriptscriptstyle (\ $}}}}

\newcommand{\qed}{\hfill$\Box$}
\newcommand{\proof}{\noindent{\bf Proof:}\ }

\newcommand{\remark}{\noindent{\bf Remark:}\ }

\newtheorem{Theorem}   {Theorem}   [section]
\newtheorem{Corollary} [Theorem]   {Corollary}
\newtheorem{Lemma}     [Theorem]   {Lemma}
\newtheorem{Proposition} [Theorem] {Proposition}

\begin{document}
\title{Topology of Event Horizon}
\author{Masaru Siino\footnote{e-mail: 
msiino@tap.scphys.kyoto-u.ac.jp, 
JSPS fellow}\\
\it Department of Physics, Kyoto University\\
Kitashirakawa, Sakyoku, Kyoto 606-01, Japan}
\maketitle
\pubnumber
\begin{abstract}
The topologies of event horizons are investigated. Considering the 
existence of the endpoint of the event horizon, it cannot 
be differentiable. Then there are the new possibilities of the topology of 
the event horizon though they are excluded in smooth event horizons. 
The relation between the topology of the event horizon and the endpoint of 
it is revealed. 
A torus event horizon is caused by two-dimensional endpoints. 
One-dimensional endpoints provide the coalescence of 
spherical event horizons. Moreover, these aspects can be removed by an 
appropriate timeslicing. The result will be useful to discuss the 
stability and generality of the topology of the event horizon\cite{MS}.
\end{abstract}

\section{Introduction}
The existence of an event horizon is one of the most
characteristic concepts of general relativity. So, many authors have studied 
the properties of the event horizon.
Mathematically, the event horizon is defined as the boundary of the
 causal past of future
 null infinity\cite{HE}. Since the natural asymptotic structure of 
 spacetimes is supposed to be
  asymptotic flat, where the topology of the future null infinity is 
  $S^2\times R$,
   we naively think that the (spatial) topology of the event horizon
    will always be $S^2$.

    Simple situations arise in general stationary spacetime, 
    for which it can be shown that any event horizon must have a spherical
     topology\cite{HA}\cite{CW}. The first work dealing with the topology of 
     non-stationary black holes is due to Gannon\cite{GA}. With the 
     physically reasonable condition of asymptotic flatness, it was proved 
     that the topology  of a smooth event horizon must be either a sphere 
     or a torus (when the dominant energy condition is satisfied). Such an approach 
     has recently been extended and generalized to give stronger 
     theorems, with the assumptions of asymptotic flatness, global 
     hyperbolicity, and a suitable energy condition. 
    Friedmann, Schleich, 
     and Witt proved the  ``topological censorship'' theorem that any two 
     causal curves extending from 
     past to the future null infinity is homotopy equivalent to each other\cite{FS}.
      Jacobson and Venkataramani\cite{JV} have 
     established a theorem that strengthens a recent result due to Browdy 
     and Galloway that the topology of an event horizon with a timeslicing is 
  a sphere if no new null generators enter the horizon at 
     later times\cite{BG}. The theorem of Jacobson and Venkataramani 
     limits the time for which a torus event horizon can persist.
    
Most of these works are based on the differentiability of the event horizon. Considering the 
whole structures of the event horizon, however, the event horizon cannot always be 
differentiable. For example, even in the case of spherically symmetric 
spacetime (for example, Oppenheimer-Snyder spacetime) the event horizon is 
not differentiable where the event horizon is formed. In general, if the 
event horizon is not smooth, we cannot say that the event horizon should be a sphere.
    
In fact, the existence of the event horizon whose topology is not a 
single $S^2$ is 
reported in the numerical 
simulations of gravitational collapses. Shapiro, Teukolsky et. al.\cite{ST}
numerically observed a torus event horizon in the collapse of a toroidal 
matter. Seidel et. al. numerically shown the 
coalescence 
of two spherical event horizons\cite{NCSA}. 
For, as shown in the present article, an event horizon is not 
differentiable at the endpoint of the null geodesic generating the event 
horizon. In the 
present  article, such an indifferentiability at the endpoint is mainly 
handled. On the contrary, we do  not care about the indifferentiability 
not related to the endpoint (for example, the indifferentiability caused 
by the pathological structure of the null infinity\cite{CG}).

In a physically realistic gravitational collapse, it is believed that 
a spacetime is quasi-stationary far in the future. So, it may be natural to 
assume that the topology of an event horizon should be a sphere for a single
 asymptotic region.
Then the problem of the topology of the event horizon is regarded as 
a topology changing process from a non-spherical 
surface to a sphere in a three-dimensional manifold (the event horizon). Therefore we will put the theories of the topology change\cite{SR}\cite{GR}
 into this problem.

In the next section, we prepare the theories of the topology 
change of a spacetime, which is applied to the event horizon in the section 3. Final section 
is devoted to the summary and discussions.
\section{The topology change of (2+1)-spacetime}
Many works have concerned the topology change of a spacetime. Some of 
these are useful to discuss the topology of an event horizon (EH) which
is a three-dimensional null surface imbedded into a four-dimensional
spacetime.
 Now we 
briefly present several theorems about the topology change of the spacetime.
\subsection{Poincar\'{e} Hopf Theorem}
Our investigation is based on a well known theorem about the relation 
between the topology of a manifold and a vector field on it\footnote{It should 
be noted that we never take the affine parametrization of a vector field 
so that the vector field is continuous even at the endpoint of the curve tangent 
to the vector field since we deal with the endpoint as the zero of the vector 
field. If we chose the affine parameters, the 
vector field would not become unique at the endpoint.}.
The following
Poincar\'{e}-Hopf theorem (Milnor 1965) is essential for our investigation.
\begin{Theorem}{\bf Poincar\'{e}-Hopf}\label{TP}
        Let $M$ be a compact $n$-dimensional $(n\ge 2)$ $C^r(r\ge1)$ manifold. 
        $X$ is 
        any $C^{r-1}$ vector field with at most a finite number of zeros, 
        satisfying following two 
        conditions. (a)The zeros of $X$ are contained 
        in $Int M$. (b)$X$ has outward directions at $\partial M$. 
        Then the sum of the indices of $X$ at all its zeros is equal to the 
        Euler number $\chi$ 
        of $M$;
        \begin{equation}
                        \chi(M)=ind(X).
                \label{}
        \end{equation}
\end{Theorem}

The index of the vector field $X$ at a zero $p$ is defined as follows. 
Let $X_a(x)$ be the components of $X$ with respect to local coordinates 
$\{x^a\}$ in a neighborhood about $p$. Set $v_a(x)=X_a(x)/\vert X\vert$. If 
we evaluate $v$ on a small sphere centered at $x(p)$, we can regard 
$v_a(S^{n-1})$ as a continuous mapping\footnote{For the theorem in this
 statement, we only need a 
continuous vector field and the index of its zero defined by the 
continuous map $v:S^{n-1}\rightarrow S^{n-1}. $Nevertheless, if one want to 
relate the index and the Hesse matrix $H=\nabla_av_b$, a $C^2$ manifold and 
a $C^1$ vector field will be required} from $S^{n-1}$ into 
$S^{n-1}$. 
The mapping degree of this map is called the index of $X$ at the zero $p$.
For example, if the map is homeomorphic, the mapping degree of the orientation 
preserving 
(reversing) map is $+1$ ($-1$). Fig.\ref{fig:zeros} gives some examples of the zeros in two 
dimensions and three dimensions.

In the present article, we treat three-dimensional manifold imbedded into 
a four-dimensional spacetime manifold as an EH. The three-dimensional manifold 
has two two-dimensional boundaries as an initial 
boundary and a final boundary (which is assumed to be a sphere in the next
 section). For such a manifold, we use the following 
modification of the Poincar\'{e}-Hopf theorem. 
Now we consider odd-dimensional manifold with two boundaries 
$\Sigma_{1,2}$. 

\begin{Theorem}{\bf Sorkin 1986}\label{TS}
                Let $M$ be a compact {\it n}-dimensional ($n > 2$ is 
                an odd number) $C^r(r\ge 1)$ manifold 
                with $\Sigma_1\cup\Sigma_2=\partial M$ and 
                $\Sigma_1\cap\Sigma_2=\phi$. $X$ is 
        any $C^{r-1}$ vector field with at most a finite number of zeros, 
        satisfying following two 
        conditions  (a)The zeros of $X$ are contained 
        in $Int M$. (b)$X$ has inward directions at $\Sigma_1$ and outward 
        directions at $\Sigma_2$. 
        Then the sum of the indices of $X$ at all its zeros is related to 
        the Euler numbers of $\Sigma_1$ and $\Sigma_2$;
        \begin{equation}
                \chi(\Sigma_{2})-\chi(\Sigma_{1})=2 ind(X).
                \label{eqn:ss}
        \end{equation}
\end{Theorem}

Its proof is given in Sorkin's work\cite{SR}.

\subsection{Geroch's Theorem}
Geroch stressed that no closed timelike curve
in a $C^r(r\ge 1)$ spacetime $(M,g)$ needs $C^{r-1}$-diffeomorphic initial and final
hypersurfaces\footnote{Originally he assumed a $C^{\infty}$-differentiable 
spacetime. Nevertheless, his theorem is easily applicable to a $C^r$ 
spacetime.}\cite{GR}.

\begin{Theorem}{\bf Geroch 1967}\label{TG}
Let $M$ be a $C^r(r\ge 1)$ $n$-dimensional compact spacetime manifold whose boundary is the 
disjoint union of two compact spacelike $(n-1)$-manifolds, $\Sigma_1$ and 
$\Sigma_2$. Suppose $M$ is isochronous, and has no closed timelike curve. 
Then $\Sigma_1$ and $\Sigma_2$ are $C^{r-1}$-diffeomorphic, and $M$ is 
topologically $\Sigma_1\times [0,1]$. 
\end{Theorem}

This theorem is not directly applicable to a null surface $H$, where
 a chronology is determined by
null geodesics generated by a null vector field $K$. In this case, ``isochronous''
means that there is no zero of $K$ in the interior of $H$. On the other hand, the closed
 timelike curve
does not rigorously correspond to a closed null curve, since on a null surface
an imprisoned null geodesic cannot be distorted, remaining
null, so as to
 become a closed curve as the theorem\ref{TG}\cite{GR}.
Then we require the strongly causal condition\cite{WA} to a spacetime
 rather than the condition of no closed causal
curve. The following modified version of Geroch's theorem
arises.
\begin{Theorem}\label{TG2}
Let $H$ be a $C^r (r\ge 1)$ $n$-dimensional compact null surface whose boundary
is the disjoint union of two compact spacelike $(n-1)$-manifolds,
$\Sigma_1$ and $\Sigma_2$. Suppose that there exists a $C^{r-1}$ null vector
 field $K$ which is
nowhere zero in the interior of $H$ and has inward and outward directions at $\Sigma_1$ and
$\Sigma_2$, respectively, and $H$ is 
imbedded into a strongly causal spacetime $(M,g)$. Then $\Sigma_1$ and
$\Sigma_2$ are $C^{r-1}$-diffeomorphic, and $H$ is topologically
$\Sigma_1\times [0,1]$.
\end{Theorem}
\proof
Let $\gamma$ be a curve in $H$, beginning
on $\Sigma_1$, and everywhere tangent to $K$. Suppose first that
$\gamma$ has no future endpoint both in the interior of $H$ and its
boundary $\Sigma_2$. Parametrizing $\gamma$ by a continuous variable $t$ 
with range zero to infinity, the infinite sequence $P_i=\gamma(i),
\ i=1,2,3,...$, on the compact set $H$ has a limit point $P$. Then for any 
positive number $s$, there must be a $t>s$ with $\gamma(t)$ in the
sufficiently small open neighborhood ${\cal U}_P$ (since $P$ is a limit
point of $P_i$), and a $t' > s$ with $\gamma(t')$ not in ${\cal
  U}_P$ (since $\gamma$ has no future endpoint). That is, $\gamma$ must
pass into and then out of the neighborhood ${\cal U}_P$ an infinite
number of times. Since ${\cal U}_P$ can be regarded as the open
neighborhood of $\gamma(t)\in {\cal U}_P$, this possibility is
excluded by the hypothesis that $H$ is imbedded into a strongly causal
spacetime $(M,g)$. Then such a curve $\gamma$ must have a future endpoint
on $\Sigma_2$ because there is no zero of $K$ which is the future endpoint of 
$\gamma$, in the interior of
$H$ from the assumption of the theorem. Hence we can draw the curve $\gamma$ through each point $p$ of
$H$ from $\Sigma_1$ to $\Sigma_2$. By defining the appropriate parameter 
of each $\gamma$, the one
parameter family of surfaces from $\Sigma_1$ to $\Sigma_2$ passing
thorough every point of $H$ is given\cite{GR}. Furthermore the $C^{r-1}$-congruence $K$
provides a one-one correspondence between any two surfaces of this
family. Hence, $\Sigma_1$ and $\Sigma_2$ is $C^{r-1}$-diffeomorphic and
$H\sim\Sigma_1\times [0,1]$.\qed


\section{The topology of event horizon}
Now we apply the topology change theories given in the previous section to 
EHs.
Let $(M,g)$ be a four-dimensional $C^\infty$ spacetime whose topology is 
$R^4$.
In the rest of this article, the spacetime  $(M,g)$ is supposed to 
be strongly causal. Furthermore, 
for simplicity, 
the topology of the EH
(TOEH\footnote{The TOEH means the topology of the spatial section of the EH. 
Of course, it
depends on a timeslicing.}) is assumed to be a smooth $S^2$
far in the future and the EH is not eternal one (in other words, the EH 
begins somewhere in the spacetime, and is open to the infinity in the 
future direction with a smooth $S^2$ section).
These assumptions will be valid when we consider only one regular 
($\sim R\times S^2$)
asymptotic region, namely the future null infinity 
$\scri^+$, to define the EH,
and the formation of a black hole. 
The following
investigation, however, will be easily extended to the case of the different final 
TOEHs far in the future.

In our investigation, the most important concept is the existence of 
the endpoints of
 null geodesics $\lambda$ which completely lie in the EH and generate it. We call them 
 the endpoints of the EH.
To generate the EH the null geodesics $\lambda$ are maximally extended to 
the future and past as long as they belong to the EH. Then the endpoint 
is the point where such null geodesics are about to come into the EH (go 
out from the EH), though the null 
geodesic  can continue to the outside or the inside of the EH through 
the    endpoint in the sense of the whole spacetime.
 We consider a null vector field $K$ on the EH which is tangent to the null 
 geodesics $\lambda$.
$K$ is not affinely parametrized but parametrized so as to be continuous 
even on the endpoint where the caustic of $\lambda$ appears. Then the endpoints of $\lambda$ are the zeros of 
$K$, which can become only past endpoints since $\lambda$ must reach to 
infinity in the future direction. Of course, taking affine parametrization, $K$ would not become 
well-defined at the endpoint. 

Moreover, the fact that the EH defined by 
$\dot{J^-}(\scri^+)$ (the boundary of the causal past of the future null 
infinity) is an achronal boundary (the boundary of a future set) tells 
us that the EH is 
an imbedded $C^{1-}$ submanifold without a boundary (see\cite{HE}).
Introducing the normal coordinates $(x^1,x^2,x^3,x^4)$ in a neighborhood 
${\cal U}_\alpha$ about $p$ on the EH, the EH is immersed as 
$x^4=F(x^1,x^2,x^3)$, where $\partial/\partial x^4$ is timelike. Since the 
EH is an achronal boundary, $F$ is a Lipschitz function and one-one map 
$\psi_\alpha:{\cal V}_\alpha\rightarrow R^3,\ \psi_\alpha(p)=x^i(p)$ is a 
homeomorphism, where ${\cal V}_\alpha$ is the intersection of ${\cal 
U}_\alpha$ and the EH\cite{HE}. Then the EH is an imbedded three-dimensional 
$C^{1-}$ submanifold.

First we pay attention to the relation between 
the endpoint and the differentiability of the EH. We see that
the EH is not differentiable at the past endpoint.    

\begin{Lemma}\label{L1}
Suppose that $H$ is a three-dimensional null surface imbedded into the spacetime $(M,g)$ by 
a function $F$ as
\begin{equation}
H:x^4=F(x^i,i=1,2,3),
\end{equation}
in a coordinate neighborhood 
$({\cal U}_{\alpha},\phi_\alpha),\ \phi_\alpha:{\cal U}_{\alpha}\rightarrow 
R^4$, where 
$\partial/\partial x^4$ is timelike.
When $H$ is generated by the set of null geodesics  whose tangent vector 
field is $K$,
$H$ and the imbedding function $F$ is indifferentiable at the endpoint of the null
 geodesic (the zero of $K$).
\end{Lemma}
\proof
If $H$ is a $C^r(r\ge 1)$ null surface around $p$, we can define the tangent space 
$T_p$ of $H$, which is spanned by one null vector and two independent 
spacelike vectors. On the contrary, there is no non-zero null vector 
$K\vert_p$ at the endpoint of the null 
geodesic generated by $K$ since the null geodesics have been extended as 
long as possible in $H$. For, the 
non-zero tangent vector will extend the geodesics and $H$ further. If one 
reparametrize $K$ so as not to vanish at the endpoint, it will become 
ill-defined. 
Then $H$ and $F$ cannot be 
differentiable at the endpoint $p$.\qed

In the present article, we just deal with this indifferentiability. So, 
we assume that the EH is $C^r(r\ge 1)$-differentiable except on the 
endpoint of the null geodesics generating the EH and the set of the 
endpoints is compact. Thus we suppose that the EH is indifferentiable 
only on compact subset.
Incidentally, in the case where the future null infinity possesses pathological
structure, the EH could be nowhere differentiable\cite{CG}. 
Nevertheless we have no concrete example of a physically reasonable spacetime 
with such a non-compact indifferentiability. Similarly there might be
the case where indifferentiable point is not the endpoint of the EH. The reason
why, in spite of this possibility, we
 consider the indifferentiability
only  caused by the endpoints, is that every EH possesses at least one
 endpoint 
except for eternal EHs. Most of the indifferentiability, which we can 
imagine, would be concerned by the endpoint. 

Next, we prepare a basic proposition.
Suppose there is no past endpoint of null geodesic generator of an EH between 
$\Sigma_1$ and $\Sigma_2$. 
 Then, Geroch's theorem stresses  the topology of the smooth
EH does not change. 

\begin{Proposition}  \label{P1}
Let $H$ be the compact subset of the EH of $(M,g)$, whose boundaries are an initial spatial section
$\Sigma_1$ and a final spatial section $\Sigma_2$,
$\Sigma_1\cap\Sigma_2=
\emptyset$. $\Sigma_2$ is assumed to be
far in the future and a smooth sphere. Suppose that $H$ is $C^r(r\ge 
1)$-differentiable. Then
the topology of $\Sigma_1$ is $S^2$.
 \end{Proposition}
\proof
If there is any endpoint of the null geodesic generator of the EH in the interior of $H$,
 $H$ cannot be $C^1$-differentiable there. Using theorem
\ref{TG2}, it is concluded that $\Sigma_1$ is topologically $S^2$, since
$H$ is imbedded into a strongly causal spacetime $(M,g)$.
\qed

Now we discuss the possibilities of non-spherical topologies.
From Sorkin's theorem there should be any zero of null vector field
 $K$ in the interior of $H$ provided that
the Euler number of $\Sigma_1
$ is different from that of $\Sigma_2\sim S^2$. Such a zero  can only be the past endpoint of the EH since the null 
geodesic generator of the EH
cannot have a future endpoint. About this past
 endpoint of the EH we state 
following two propositions.

\begin{figure}
\centerline{\epsfxsize=18cm\epsfbox{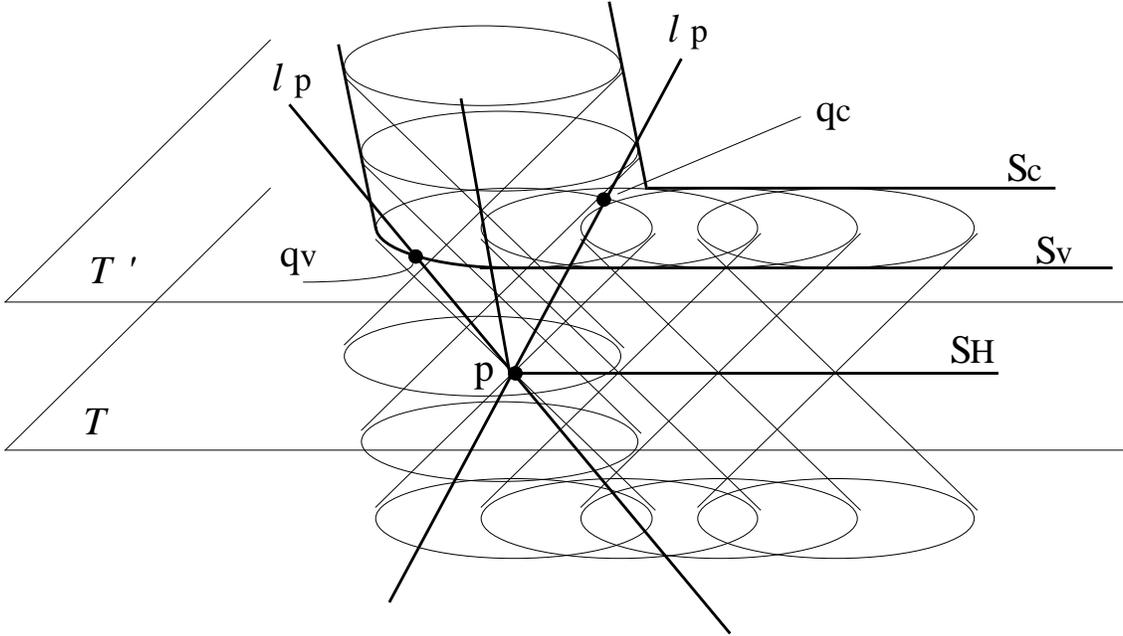}}
\caption{The neighborhood of $p$ is sliced by two spatial hypersurfaces
  $\cal T$ and $\cal T'$.
$S_H$ is on the lower hypersurface $\cal T$. $\ell_p$ passes through $p$. 
In the convex (concave) case, the EH is given by the enveloping surface 
$S_v$ ($S_c$). $q_v$ ($q_c$) is a point on $\ell_p$ at a little future of $p$. 
$S_v$ is $C^1$-differentiable at $q_v$. $q_c$ is inside $S_c$.}
\label{fig:cones}
\end{figure}

\begin{Proposition}\label{P2}  
           The set of the past endpoints 
 of the EH 
is a spacelike set.
\end{Proposition}
\proof
The set of endpoints (SOEP) is obviously an achronal set as the EH is a 
null surface (achronal boundary). 
Suppose that the SOEP includes a null segment $\ell_p$ through an event $p$. By the 
lemma \ref{L1}, the null segment 
$\ell_p$ is the indifferentiable points of the EH. The EH, however, is differentiable in the null direction tangent to 
$\ell_p$ at $p$ since $\ell_p$ is a smoothly imbedded into the smooth 
spacetime $(M,g)$. Then the
 section $S_H$ of the EH on a spatial hypersurface through 
$p$ is indifferentiable at $p$ as shown in Fig.\ref{fig:cones}. 
Considering a sufficiently small neighborhood ${\cal U}_p$ about $p$, 
the local causal
structure of ${\cal U}_p$ is 
similar to 
that of Minkowski spacetime, since $(M,g)$ is smooth there. Therefore, when $S_H$ is 
convex at $p$, the EH will be $C^1$-differentiable at $q_v$ which is on a little future of the 
null segment $\ell_p$ (see Fig.\ref{fig:cones}) because the EH is the outer side of the 
enveloping surface 
of the 
light cones standing along $S_H$ in the neighborhood ${\cal U}_p$ about $p$. 
Nevertheless, from the lemma \ref{L1}, also the endpoint $q_v$ cannot be 
smooth in this section.  On the contrary, if $S_H$ is concave, $q_c$ which is on a little 
future of $\ell_p$ will invade the inside of the EH (see 
Fig.\ref{fig:cones}). Thus the SOEP 
cannot contain either convex and concave null segment. 
Moreover if two disconnected segments could be connected by a null geodesic, 
the future endpoint of the null geodesic generator would exist. Hence the SOEP is spacelike set.
\qed

\begin{Proposition} \label{P3}
           The SOEP of the EH of $(M,g)$ is arc-wise connected. Moreover, the collared SOEP is 
           topologically $D^3$.
\end{Proposition}
\proof
Consider all the null geodesics $\lambda_{p_e}(\tau)$ emanating from the 
SOEP $\{p_e\}$ tangent to the null 
vector field $K$. Since the SOEP is the set of the zeros of $K$, $p_e$ 
corresponds to $\lambda_{p_e}(-\infty)$. From the proposition \ref{P2}, 
the spacelike section 
$\cal S$ of the EH very close to the SOEP $\{p_e\}$, is determined by a 
map $\phi^K$, with a small parameter $\Delta \tau$ of the null 
geodesic $\lambda_{p_e}$;
 \begin{eqnarray}
        \phi^K :\{q\in{\cal S}\}&\longrightarrow& \{p_e\}\\
        s.t.\  \lambda_{p_e}(-\infty)&=&p_e,\ \lambda_{p_e}(\Delta\tau)=q.
        \label{}
 \end{eqnarray}
Here, with a sufficiently small $\Delta\tau$ ($\rightarrow -\infty$), $K$ has inward directions
 to $H$ at $\cal S$, 
where $H$
is the subset of the EH bounded by $\cal S$ and the final spatial section $\Sigma_2$ 
which is far in the future and a smooth sphere from the assumption.
By this construction, all the endpoints are wrapped by $\cal S$ and $\cal 
S$ is compact because of the assumption that the SOEP is compact. 
$H$ and the SOEP are on the opposite side of $\cal S$.
 Therefore there is no endpoint in the interior of $H$. Since $H$ is 
 $C^r(r\ge 1)$-differentiable except on the SOEP and compact from the 
 assumption, the proposition \ref{P1} implies that $\cal S$ is homeomorphic to $ 
 \Sigma_2\sim S^2$ and $H$ is topologically $S^2\times [0,1]$. 
 If there were two or more connected components of the SOEP, one would 
need the same number of spheres to wrap it with $\cal S$ being sufficiently close to the 
SOEP. However, since $\cal S$ is homeomorphic to a single $S^2$, the SOEP 
should be 
arc-wise connected. In other wards, the collared SOEP is topologically
 $D^3$, because the EH and the SOEP are imbedded into $(M,g)$.\qed

Now we give theorems and corollaries about the topology of the 
spatial section of the EH on a timeslicing.
First we consider the case where the EH has simple structure.

\begin{Theorem}\label{T1}
Let $S_H$ be the section of an EH 
by a spacelike hypersurface.
If the EH is $C^r(r\ge 1)$-differentiable at $S_H$, it is
topologically $\emptyset$ or $S^2$.
\end{Theorem}
\proof
From the proposition \ref{L1}, there is no endpoint of the EH on $S_H$. Since the EH
is assumed not to be eternal, there exists at least one endpoint of the EH in the past of $S_H$
as long as $S_H \neq \emptyset$. Therefore the proposition \ref{P3} implies
there is no endpoint of the EH in the future of $S_H$. By the  
assumption that the EH is $C^r(r\ge 1)$-differentiable except on the SOEP and the proposition
\ref{P1}, it is concluded that $S_H$ is topologically $S^2$.
\qed

On the other hand, we get the following theorem about the change of the  
 TOEH with the aid of Sorkin's
 theorem. 
\begin{Theorem}\label{T2}
Consider a smooth timeslicing ${\cal T}={\cal T}(T)$ defined by a smooth 
function $T(p)$;
\begin{equation}
        {\cal T}(T)=\{p\in M\vert T(p)=T=const.,\ \  T\in 
        \left[T_1,T_2\right]\}, \ \ g(\partial_T,\partial_T)<0.
        \label{}
\end{equation}
Let $H$ be the subset of the EH cut by ${\cal T}(T_1)$ and ${\cal T}(T_2)$, 
whose boundaries are the initial spatial section $\Sigma_1\subset 
{\cal T}(T_1)$ and the final spatial section $\Sigma_2\subset 
{\cal T}(T_2)$, and $K$ be the null vector field generating the EH.
 Suppose that $\Sigma_2$ is a 
sphere.  If, 
in the timeslicing $\cal T$, the TOEH changes ($\Sigma_1$ is not
homeomorphic to $\Sigma_2$) then there is the SOEP (the zeros of $K$) in 
$H$, and
\begin{itemize}
        \item the one-dimensional segment of the SOEP causes the coalescence of two
        spherical EHs.
        \item  the two-dimensional segment of the SOEP causes the change of the TOEH
        from a torus
        to a sphere.
\end{itemize}        
\end{Theorem}
\proof
First of all, we regularize $H$ and $K$ so that the theorem \ref{TS} can be applied 
to this case.
Introducing normal coordinates $(x^1,x^2,x^3,x^4)$ in a neighborhood 
${\cal U}_\alpha$ about $q\in H$,  since 
the EH is an achronal boundary,. 
$H$ is imbedded by a Lipschitz function $x^4=F(x^i,i=1,2,3)$, where 
$\partial/\partial x^4$ is timelike 
(see\cite{HE}). Here we set $x^4(p)=T(p)-T(q)$ in ${\cal U}_\alpha$ 
about $q$. Since $M$ is a metric space, there is the partition of unit 
$f_\alpha$ for the atlas 
$\{{\cal U}_\alpha,\phi_\alpha\},\  
\phi_\alpha:{\cal U}_\alpha\rightarrow R^4$\cite{SS}. Then a smoothed function of 
the Lipschitz function $T(p\in H)$ (which is restricted on indifferentiable 
submanifold $H$ for the smooth function $T(p)$ to become indifferentiable) with a smoothing scale 
$\epsilon$ is given by 
\begin{eqnarray*}
        \widetilde{T}(p\in H)&=&\Sigma_\alpha\int_{\cal{U}_\alpha}f_\alpha T(r=\phi_\alpha^{-1}(x^1,x^2,x^3,x^4)) 
        W(p,r)\delta(x^4-F(x^1,x^2,x^3)) dx^1dx^2dx^3dx^4\\
        &=&\Sigma_\alpha\int_{\cal{U}_\alpha}f_\alpha (F(x^1,x^2,x^3)+T(q)) 
        W(p,r)\delta(x^4-F(x^1,x^2,x^3)) dx^1dx^2dx^3dx^4\\
        W(p,r)&=&0,\ \ \ p\notin {\cal U}_\alpha \\
        W(p,r)&=&w(     \vert p-r \vert), \ \ \ p\in {\cal U}_\alpha\\
        \vert p-r \vert&=& 
        \sqrt{(x^1_p-x^1_r)^2+(x^2_p-x^2_r)^2+(x^3_p-x^3_r)^2+(x^4_p-x^4_r)^2} \ \ \ in\ \ {\cal U}_\alpha
         \ \ about\ \ q\\
        w(x)&\le&\infty,\ \ \ w(x\gg\epsilon)\ll 1,\ \ \ \int w(x)=1\ \ \ ,
\end{eqnarray*}
where $w$ is an appropriate window function with a smoothing scale $\epsilon$.
The support of $W$ is a sphere with its radii $\sim \epsilon$ and 
$w(\vert x\vert,\epsilon\rightarrow 0)= \delta^4({\bf x})$. Of course, $\epsilon =0$ gives 
the original function $T(p\in H)=\widetilde{T}(p\in H)$. Taking sufficiently small 
non-vanishing $\epsilon$, a new imbedded submanifold 
$\widetilde{H}$, with $\widetilde{x^4}(p\in \widetilde{H})=\widetilde{T}(p)-
\widetilde{T}(q)=:\widetilde{F}(x^1,x^2,x^3)$ in ${\cal U}_\alpha$ about 
$q$, can 
become homeomorphic to $H$ and $C^r(r\ge 1)$-differentiable. From this 
smoothing procedure, we define a smoothing map $\pi$ (homeomorphism);
\begin{eqnarray}
        \pi : H&\longrightarrow& \widetilde{H}\\
\phi_\alpha^{-1}(x^1,x^2,x^3,x^4)&\longrightarrow&\phi_\alpha^{-1}(x^1,x^2,x^3,\widetilde{x^4}).
        \label{}
\end{eqnarray}
Of course, this map depends on the atlas $\{{\cal 
U}_\alpha,\phi_\alpha\}$ introduced.
This smoothing map induces the following correspondences,
\begin{eqnarray}
\lambda &\longrightarrow& \widetilde{\lambda},\ \ \ 
\Sigma_{1,2}\longrightarrow \widetilde{\Sigma_{1,2}},\\ 
{\cal T} &\longrightarrow& \widetilde{\cal T},\ \ \ 
\pi^*:K\longrightarrow \widetilde{K},
\end{eqnarray}
where $\widetilde{K}$ is the tangent vector field of curves 
$\widetilde{\lambda}$ generating $\widetilde{H}$. 
Now $\widetilde{K}$ is not always null. 

\begin{figure}
\centerline{\epsfxsize=14cm \epsfbox{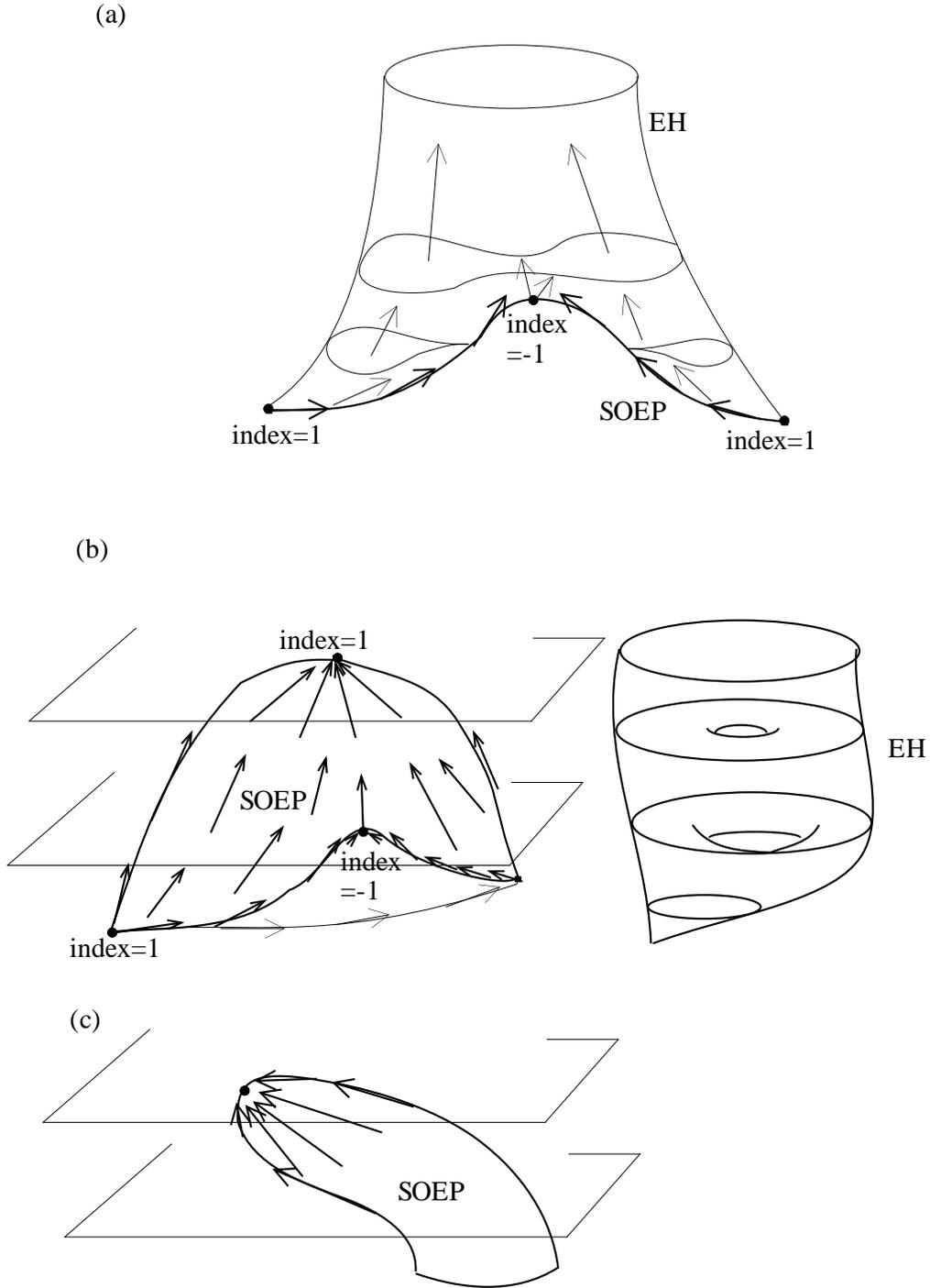}}
\caption{(a) and (b) are the one-dimensional and two-dimensional SOEP,
 respectively.  In (b), we draw the whole of the EH separately. (c) is the case
  in which the edge of the SOEP is hit from the future.
 By these vector fields $\overline{K}$, 
  the SOEPs are generated. The zeros of $\overline{K}$ and their indices are indicated.}
\label{fig:ends}
\end{figure}

Furthermore, using the transformed
timeslicing $\widetilde{\cal T}$, we should modify $\widetilde{K}$ so that the SOEP of 
$\widetilde{\lambda}$ becomes zero-dimensional set that is, the set of isolated zeros, (where the SOEP will no longer 
 always be arc-wise connected). To make the SOEP zero-dimensional, a 
 modified vector field $\overline{K}$ should 
be given on the SOEP of $\widetilde{K}$ so as to generate the SOEP. 
On the SOEP of $\widetilde{K}$, $\overline{K}$ is determined by the timeslice 
$\widetilde{\cal T}$ so that $\overline{K}$ is tangent to the SOEP of $\widetilde{K}$ and 
directed to the future in the sense of the timeslicing $\widetilde{\cal T}$. 
Especially, at the boundary of the SOEP of $\widetilde{K}$, we should be 
careful that $\overline{K}$ is tangent also to the non-zero-dimensional 
boundary of the SOEP.
 Here it is noted that the case in which the boundary is 
tangent to the timeslicing $\widetilde{\cal T}$ is possible and we cannot 
determine the direction of $\overline{K}$ there. 
Since such a situation is unstable under the small deformation of the 
timeslicing, however, we omit this possibility as mentioned in the 
remark appearing after this proof. Hence $\overline{K}$ is determined on the SOEP of 
$\widetilde{K}$ (see, for example, Fig.\ref{fig:ends}) and has some 
isolated zeros there. 
At this step, $\overline{K}$ on the SOEP and $\widetilde{K}$ except on 
the SOEP is not continuous. Then, we modify $\widetilde{K}$ around the 
SOEP along $\overline{K}$, and make modified $\widetilde{K}$ into 
$\overline{K}$ except on the SOEP without changing the characters of the 
zeros, so that $\overline{K}$ becomes 
continuous vector field on $\widetilde{H}$. One may be afraid that the extra 
zero of $\overline{K}$ appears by this continuation. Nevertheless it is 
guaranteed by the existence of the foliation by the timeslice $\cal T$ or 
$\widetilde{\cal T}$ that there exists the desirable modification of 
$\widetilde{K}$ around the SOEP, since both $\widetilde{K}$ and 
$\overline{K}$ are future directed in the sense of the timeslicing 
$\widetilde{\cal T}$. Thus we get 
$\overline{K}$ and its integral curves 
$\overline{\lambda}$ on the whole of $\widetilde{H}$. From this construction of $\overline{K}$, there are 
some isolated zeros of $\overline{K}$ only on the SOEP of $\widetilde{K}$ and $\overline{K}$ is everywhere 
future directed in the sense of the timeslicing $\widetilde{\cal T}$ 
(though they will be spacelike somewhere). Of course, $\overline{\lambda}$ 
will have both future and past endpoints.

Now we apply the theorem \ref{TS} to $\widetilde{H}$ with the 
modified vector field $\overline{K}$, whose boundaries are 
$\widetilde{\Sigma_1}$ and $\widetilde{\Sigma_2}\sim S^2$. Since $\widetilde{\Sigma_1}$ and 
$\widetilde{\Sigma_2}$ are on ${\cal T}(T_1)$ and ${\cal T}(T_2)$, 
respectively, $\overline{K}$ 
has inward directions at $\widetilde{\Sigma_1}$ and outward directions at 
$\widetilde{\Sigma_2}$. 

\begin{figure}
        \centerline{\epsfxsize=13cm \epsfbox{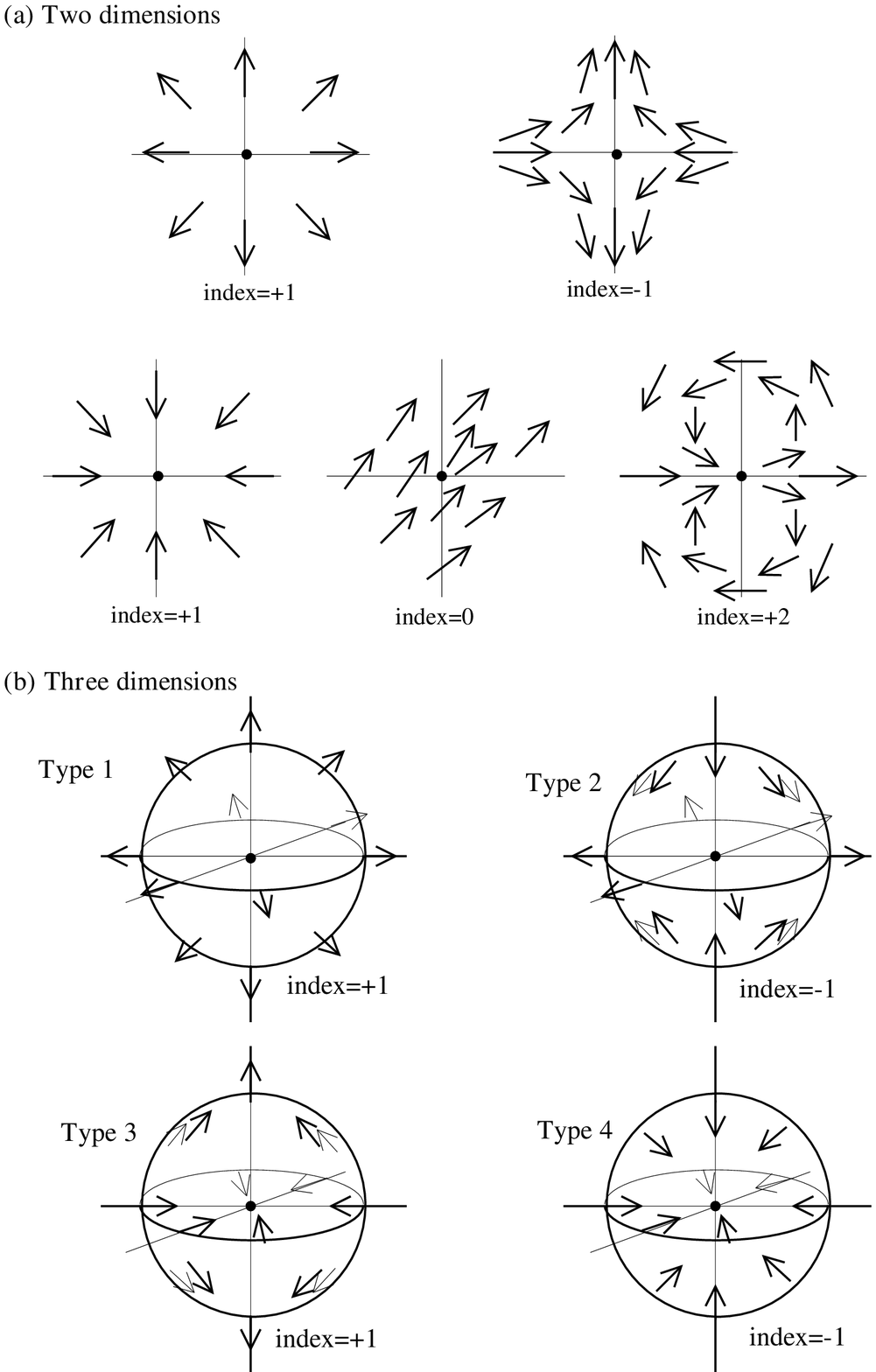}}
        \caption{(a) Two-dimensional zeros and a
 vector field around them are shown.
        Five types of the zeros are shown in this figure.
        (b)Three-dimensional zeros and a vector field around them are shown. 
        Only the zeros with $|\rm index |=1$ are shown.
 Other cases can easily
        be known by the analogy of (a).}
        \protect\label{fig:zeros}
\end{figure}

 From the construction above, we see that the type of the zero of $\overline{K}$ depends
on the dimensions of the SOEP. Especially, for the zero most 
in the future, the one-dimensional SOEP
provides the zero of the second type in Fig.\ref{fig:zeros}(b) corresponding 
to index $=-1$ and the two-dimensional 
SOEP gives that of the third type in Fig.\ref{fig:zeros}(b) with index $=+1$ (see 
Fig.\ref{fig:ends}). Following the theorem 
\ref{TS}, the Euler number
changes at the zero by $2\times$index. Therefore if there is the
one- (two)-dimensional SOEP, the timeslicing $\cal T$ gives the topology change 
of the EH from two spheres (a torus) to a sphere. When $H$ 
contains the whole of the SOEP, it will, according to the theorem \ref{TS}, present all changes 
of the TOEH from the formation of the EH to a sphere far in the future as shown in 
Fig.\ref{fig:ends}. To complete discussions, 
we also consider dull cases provided 
by a certain timeslicing. When the edge of the SOEP is hit by the  
timeslicing from the future, according to the 
construction above, it gives a zero with its index being zero 
(Fig.\ref{fig:ends}(c)) and there is no topology change of the EH. \qed

This result is partially suggested in Shapiro, Teukolsky, and Winicour\cite{ST}.

\begin{figure}
\centerline{\epsfxsize=18cm \epsfbox{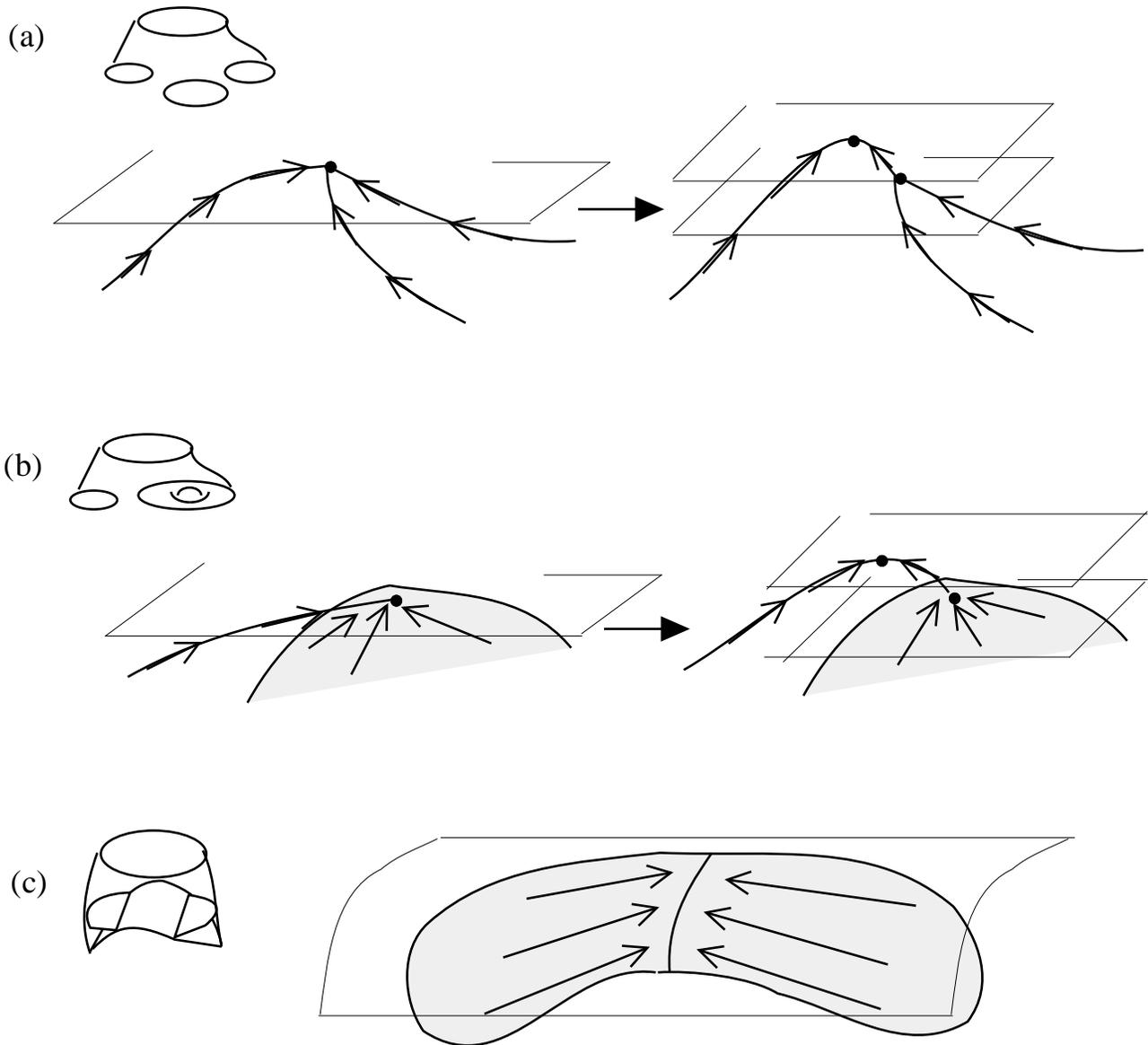}}
\caption{(a) and (b) are the example of the branching SOEP 
in an accidental timeslicing. They are understood by the small 
deformation of the timeslicing. 
On the other hand, (c) is the case in which the timeslicing is 
partially tangent to the SOEP.
The two-dimensional SOEP behaves as the one-dimensional SOEP.}
\label{fig:branches}
\end{figure}

\remark
One may face special situations.
 The possibilities of the branching 
endpoints should be noticed. If the SOEP
possesses a branching point, a special timeslicing can make the branching 
point into an isolated zero though such a timeslicing loses this aspect under 
the small deformation of the timeslicing. The index
of this branching endpoint may deny a direct consideration.
 The situation, however, is regarded as the degeneration of the two 
 distinguished zeros of $\overline{K}$ in  
 $\widetilde{H}$. Some of examples are displayed in Fig.\ref{fig:branches}.
 Imagine a little slanted timeslicing, and it will decompose the branching point
 into two distinguished (of course, there are the possibilities of the degeneration of three or more)
zeros. The first case is the branch of
  the one-dimensional SOEP\footnote{We can also
   treat the branching   points of the two-dimensional SOEP in the same
   manner.} (Fig.\ref{fig:branches}(a)), where
  the branching point is the degeneration of two zeros of $\overline{K}$ 
  with their index
   being a minus one, since they are the results of 
   the one-dimensional 
   SOEP. Then the index of the branching point is a minus two and, for example, three
   spheres coalesce there. The next case is a one-dimensional branch from
    the two-dimensional SOEP (Fig.\ref{fig:branches}(b)).
This branching point is the degeneration of the zeros of $\overline{K}$ from
 the one-dimensional
SOEP (index $=-1$) and the two-dimensional SOEP (index $=+1$). 
This decomposition tells that though the index of this point vanishes, the TOEH changes at
 this point, for example, from a sphere and a torus to a sphere. 
 Of course, the Euler number does not change in this process. 
 Furthermore, these topology changing processes are
 stable under the small deformation of the timeslicing. 
 Finally, there is the case in which a timeslicing is partially tangent to the SOEP 
 or its boundary.
 For instance, an accidental timeslicing can hit not a single point of 
 the SOEP but a line of the SOEP from the future 
 as shown in Fig.\ref{fig:branches}(c). For such a timeslicing, the contribution of 
 the two-dimensional SOEP to the index 
 is not a minus one but one. This situation, however, is
  unstable under
  the small deformation of the timeslicing, and we omit such a case in the 
  following.

\begin{figure}
\centerline{\epsfxsize=18cm \epsfbox{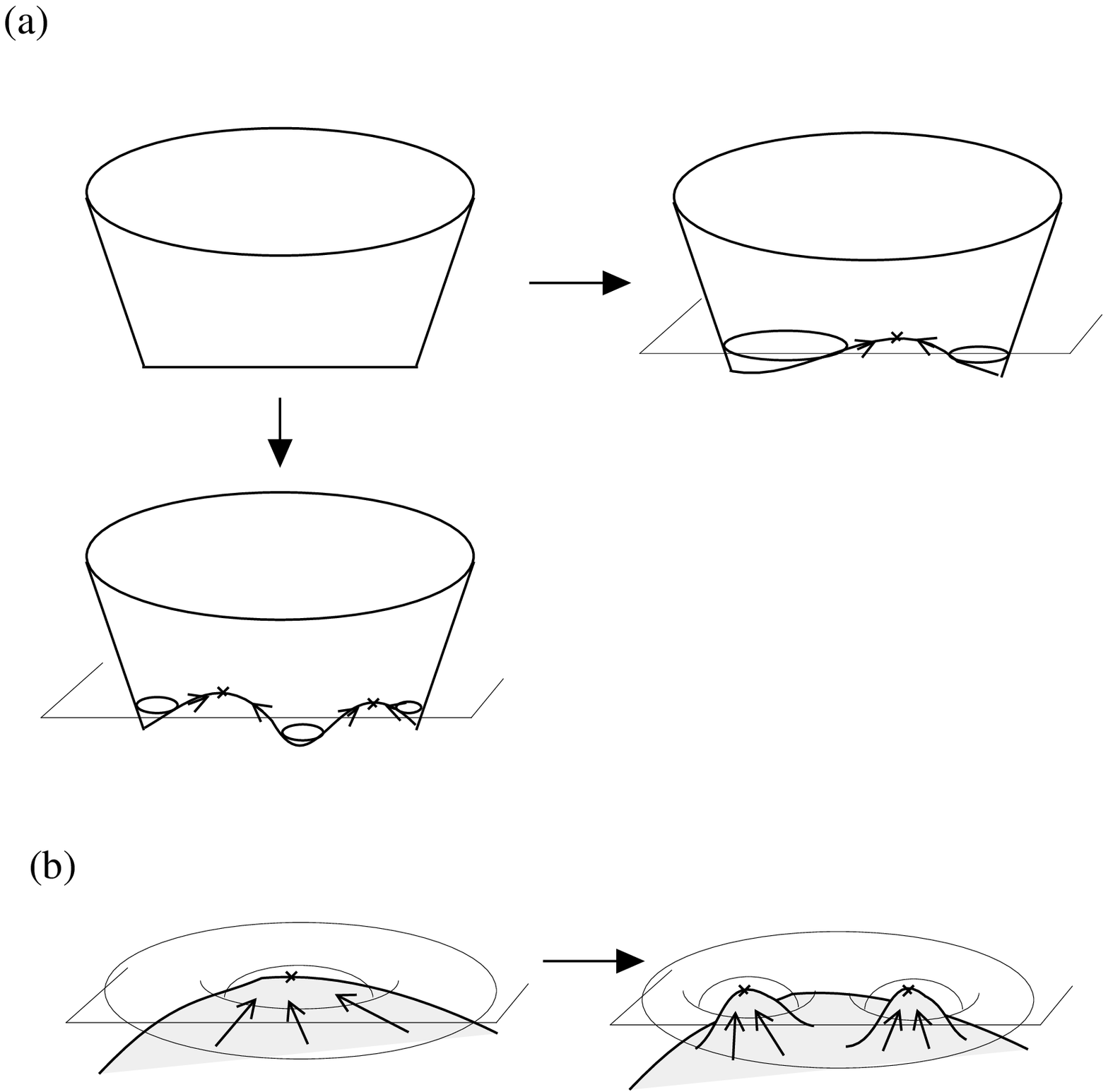}}
\caption{By the appropriate change of the timeslicing, 
the number of zeros of the vector field $\overline{K}$ changes. (a) and (b) are  
the one-dimensional and two-dimensional SOEP, respectively.}
\label{fig:multi}
\end{figure}

Incidentally,
 a certain timeslicing gives the further changes of the Euler number. 
\begin{Corollary}\label{CL1}
The topology changing processes of an EH from $n\times S^2$
to $S^2$, ($n=1,2,3,...$) can change each other,   and 
from a surface with genus=$n$  to $S^2$, ($n=1,2,3,...$) can also change,
  under the  appropriate
deformation of their timeslicing. 
\end{Corollary}
\proof
From the theorem \ref{T2}, when the TOEH changes from $n \times S^2$ to a 
single $S^2$ in a timeslicing, there should be the one-dimensional SOEP 
(in which there may be some branches).
 Since the SOEP is a spacelike set (the proposition \ref{P2}),
 there is another appropriate timeslicing hitting the SOEP at $m$ 
 different points simultaneously (Fig.\ref{fig:multi}(a)). On this timeslicing, the Euler number changes by $-2
   \times m$ and $m$$+$$1$ spheres coalesce. 
   In the same logic, the EH of a surface with genus=$n$ can be regarded as
   the EH of a surface with genus=$m$
   by the appropriate change of its timeslicing (see Fig.\ref{fig:multi}(b)). \qed
   
As shown in the corollary \ref{CL1}, the TOEH  highly depends on the 
timeslicing. Nevertheless, the theorem \ref{T2} tells that there is the 
distinct difference
between the coalescence of $n$ spheres  where the Euler number decreases by the
one-dimensional SOEP and the EH of a surface with genus=$n$ where the Euler number
increases by the two-dimensional SOEP.
Finally we see that, in a sense, the TOEH is a transient term.

\begin{Corollary}
          All the changes of the TOEH are reduced to the trivial creation of 
          an EH which is topologically $S^2$.
\end{Corollary}
\proof
From the proposition \ref{P2}, the collared SOEP is topologically $D^3$. 
Therefore, since the SOEP is spacelike,
 there is a certain timeslicing in which $p_c$ is most in the past on the 
 SOEP and the more distant from $p_c$, the more in the future. 
 In this timeslicing, $\widetilde{H}$ has only one significant zero $p_c$ of
 $\overline{K}$ (type 1 in Fig.\ref{fig:zeros}(b)), which corresponds to
 the point where the EH is formed and meaningless zeros (with the index 
 0, for example, see Fig.\ref{fig:ends}(c)) on the edge of the SOEP. The index of $p_c$ is +1, and a 
 spherical EH are formed there. \qed

\begin{figure}
\centerline{\epsfxsize=18cm \epsfbox{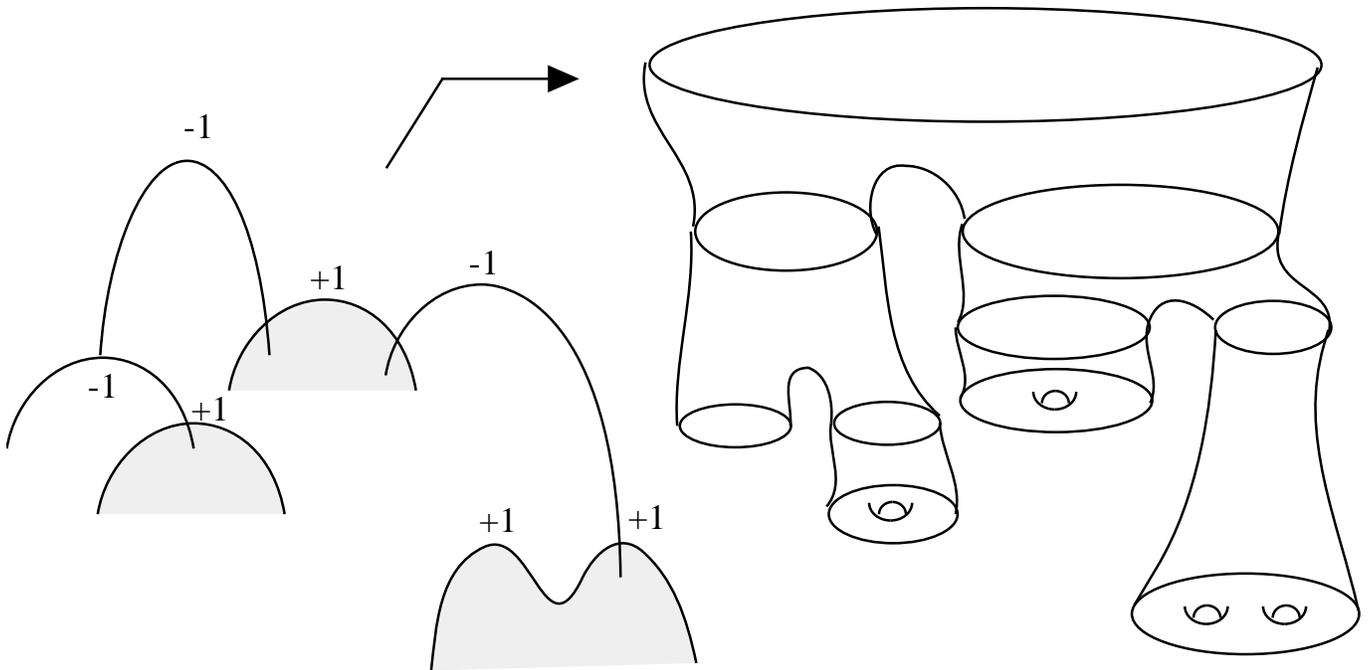}}
\caption{An example of the graph of the SOEP is drawn. Determining the order of the 
vertices, we see the TOEH from the index of each zero.}
\label{fig:graph}
\end{figure}

Thus we see that the change of the TOEH is determined by the topology of 
the SOEP and the timeslicing way of it. For example, we can imagine the graph of the SOEP as Fig.\ref{fig:graph}.
To determine the TOEH we must only give the order to each
 vertex of  the graph by a timeslicing. The graph in Fig.\ref{fig:graph}
 might be rather complex. Nevertheless, considering a small scale
 inhomogeneity, for example the scale of a single particle, the EH may admit
 such a complex SOEP. It will be smoothed out in macroscopic physics.

\section{Summary and Discussion}
We have studied the topology of the EH (TOEH), partially considering the 
indifferentiability of
the EH. We have found that the coalescence of EHs is related to the one-dimensional
 past endpoints and a torus EH two-dimensional. In a sense, this is the 
 generalization of the result of Shapiro, Teukolsky, and 
 Winicour\cite{ST}.  Furthermore these
changes of the TOEH can be removed by an appropriate timeslicing
since the set of the endpoints (SOEP) is a connected spacelike set. We see that
 the TOEH strongly
depends on the timeslicing. The dimensions of the SOEP,
however, play an important role for the TOEH and, of course, are invariant under the change of the 
timeslicing. 

Based on these results, it arises the question what 
controls the dimensions of the SOEP. One may expect that something like an
energy condition restricts the possibilities of the SOEP. Nevertheless it is doubtful 
since, in fact, two cases with the non-trivial TOEH (the coalescence of EHs---the one-dimensional SOEP 
and a torus EH---the two-dimensional SOEP), where the energy condition is 
satisfied, are reported
 in the numerical simulations \cite{ST}\cite{NCSA}. Are these generic in 
 real gravitational collapses?
It is probable that the gravitational collapse in which the EH is a 
single sphere for each timeslicing is not generic, since the zero-dimensional 
SOEP reflects the higher symmetry
of a system 
than that of the one- or two-dimensional SOEP.
On balance, the symmetry of matter 
configurations will control it. For example, it is possible to discuss 
the stability and generality of such a symmetry. We will show the 
stability of a spherical EH under linear perturbations and the 
catastrophic structure of the SOEP\cite{MS}. These discussions would tell 
something about how the structure of the SOEP is determined dynamically.

In the present article, we have assumed some conditions about the structure of 
spacetime. Can other weaker conditions take the place of them? 
First, the strongly causal condition may be too strong. For, this condition 
is needed only on the EH. 
For example, the global hyperbolicity implies the strong causality on the EH, 
because the global hyperbolicity exclude a closed causal curve and a  past 
imprisoned causal curve and there should be no future imprisoned null curve on 
the EH.
Next, we required that  
the TOEH is smooth $S^2$ far 
in the future. This, however, is not crucial. Since the present 
investigation is based on the topology change theory, the same discussion 
 is possible for other final TOEHs. 
Next, the $C^r(r\ge 1)$-differentiability of the EH is supposed except on the 
compact SOEP while 
it might be able to be violated in realistic situations. It is not clear 
whether this differentiability can be implied by other physically reasonable conditions. The 
indifferentiability, however, is overwhelmingly easier to occur on the 
endpoint than not on the endpoint. 
Every non-eternal EH possesses such an indifferentiable point as a past 
endpoint and we do not 
have any simple example where the indifferentiable point is not the endpoint.
On the other hand, the case in which the EH is not indifferentiable only 
on  compact subsets (i.e., the SOEP is not compact) would be 
excluded by the realistic requirement about the asymptotic structure of 
the spacetime, as 
a nowhere differentiable spacetime\cite{CG} is excluded by 
asymptotic flatness.
It would be worth to clarify such properties about the differentiability 
of the EH.

Incidentally, some of the statements in this article may overlap the result 
of the former works\cite{HA}$\sim$\cite{BG}. Nevertheless the condition required here is 
pretty different from that of them (for example, the energy condition 
have never been assumed here). They might be the extension of the former work. 

Finally we are reminded of an essential question. How can we see the topology of the EH?
Some of the former works, for example ``topological censorship''\cite{FS}, oppose.
On the contrary, we expect the phenomena highly depending on the existence of the EH as 
the boundary condition of fields, for 
instance the quasi-normal mode of gravitational wave\cite{CH} or Hawking 
radiation\cite{HR}, reflect the TOEH. This is our future problem.

\centerline{\bf Acknowledgements}

We would like to thank Professor H. Kodama, Dr. K. Nakao, Dr. T. Chiba, Dr. 
A. Ishibashi, and Dr. D. Ida. 
for helpful discussions. We are grateful to Professor H. Sato and 
Professor N. Sugiyama for their continuous encouragement.
The author thanks the Japan Society for the 
Promotion of Science for financial support. This work was supported in 
part 
by the Japanese Grant-in-Aid for Scientific Research Fund of the Ministry 
of 
Education, Science, Culture and Sports.

\end{document}